\documentclass[a4paper,referee]{raa}           
\usepackage{graphicx,times}
\usepackage{natbib}
\usepackage{amssymb,amsmath}
\usepackage{CJKutf8}
\bibpunct{(}{)}{;}{a}{}{,}

\usepackage{hyperref}
\hypersetup{colorlinks = true, linkcolor = green, anchorcolor = red, citecolor = blue, filecolor = red, pagecolor = red, urlcolor = red}

\begin{document}

   \title{New Constraint of the Hubble Constant by Proper Motions of Radio Components Observed in AGN Twin-jets
\footnotetext{\small $*$ Supported by the National Natural Science Foundation of China (No. 11903002), and the Research Project of Baise University (No. 2019KN04).}
}

 \volnopage{ {\bf 20XX} Vol.\ {\bf X} No. {\bf XX}, 000--000}
   \setcounter{page}{1}

   \author{
\begin{CJK*}{UTF8}{gbsn}
Wei-Jian Lu(陆伟坚)\inst{1}, Yi-Ping Qin(覃一平)\inst{2}
\end{CJK*}
   }

   \institute{School of Information Engineering, Baise University, Baise 533000, China; {\it william\_lo@qq.com}(Corresponding author)\\
        \and
Center for Astrophysics, School of Physics and Materials Science, Guangzhou University, Guangzhou 510006, China; {\it  1476426056@qq.com}(Corresponding author)\\
\vs \no
   {\small Received 2021-***; accepted 2021-7-12}
}

\abstract{As the advent of precision cosmology, the Hubble constant ($H_0$) inferred from the Lambda Cold Dark Matter fit to the Cosmic Microwave Background data is increasingly in tension with the measurements from the local distance ladder. To approach its real value, we need more independent methods to measure, or to make constraint of, the Hubble constant. In this paper, we apply a plain method, which is merely based on the Friedman-Lema\^itre-Robertson-Walker cosmology together with geometrical relations, to constrain the Hubble constant by proper motions of radio components observed in AGN twin-jets. Under the assumption that the ultimate ejection strengths in both sides of the twin-jet concerned are intrinsically the same, we obtain a lower limit of the $H_{\rm 0,min}=51.5\pm2.3\,\rm km\,s^{-1}\,Mpc^{-1}$ from the measured maximum proper motions of the radio components observed in the twin-jet of NGC 1052.
}

   \authorrunning{W.-J. Lu \& Y.-P. Qin}            
   \titlerunning{New Constraint of the Hubble Constant}  
   \maketitle

%
\section{Introduction}           
\label{sect:intro}
The Hubble constant ($H_0$) parametrize the current expansion rate of the Universe  \citep{Hubble1929}. Locally, redshifts of the nearby extragalactic sources are very small ($z\ll1$), and then within the framework of the Friedman-Lema\^itre-Robertson-Walker (FLRW) cosmology, the Hubble law can be maintained for all kinds of universe
\begin{equation}
D_A(1+z)\approx\frac{cz}{H_0},
\end{equation}
where $D_A$ is the angular distance, and $c$ is the speed of light. As one of the most important quantities in cosmology, the Hubble constant can characterize not only the current age of the Universe, but also the overall extragalactic distance scale. For decades, considerable resources have been devoted to improve precision of the Hubble constant measurements, using a variety of independent methods  (e.g., \citealp{Planck2020,Riess2019,Abbott2018,Burns2018,Freedman2019,
Huang2020,Pesce2020,Wong2020,Birrer2020,Abbott2017,
Hotokezaka2019,Wang2020}). 

Unexpectedly, as the advent of precision cosmology, the Hubble constant inferred from the early universe is increasingly in tension with the value available from local measurements. As reported recently, the Planck collaboration finds $H_0 = 67.27 \pm 0.60\,\rm km\,s^{-1}\,Mpc^{-1}$ based on the Lambda Cold Dark Matter ($\Lambda$CDM) fit to the Cosmic Microwave Background (CMB) data \citep{Planck2020}, while combining 70 long-period Cepheids in the Large Magellanic Cloud observed by the Hubble Space Telescope, the local measurement presented by the 2019 SH0ES \citep{Riess2019} collaboration obtains $H_0 = 74.03\pm1.42\,\rm km\,s^{-1}\,Mpc^{-1}$ . These two typical measurements are in tension at about 4.4$\sigma$. As being pointed out recently, this fact may indicating that we are seeing the first signs of new physics beyond the $\Lambda$CDM standard cosmological model \citep{Riess2019na} .

At this crossroad of cosmology, we urgently need more new methods to constrain the Hubble constant. The new methods would be desirable if they do not rely on either traditional distance ladders or the standard $\Lambda$CDM cosmological model. {Indeed, more and more methods of the Hubble constant measurement have been developed. It is now possible to use, for example, the gravitational-wave standard sirens (the gravitational-wave analog of astronomical standard candles, e.g., \citealp{Abbott2017,Zhao2018,Yu2020}), water megamasers residing in the accretion disks around supermassive black holes (SMBHs) in active galactic nuclei (AGNs, e.g., \citealp{Pesce2020}), strong gravitational lensing effects on quasar systems (e.g., \citealp{Wong2020}), spectroastrometry and reverberation mapping of AGN (e.g., \citealp{Wang2020}), to constrain the Hubble constant. In particular, standard sirens may play an important role in the next decade (e.g., \citealp{Abbott2017,Zhao2018,Yu2020}), since this measurement does not require traditional distance ladders and it is independent of the standard $\Lambda$CDM cosmological model. }

In this paper, we apply a quite plain method, which is merely based on the FLRW cosmology together with geometrical relations, to constrain the Hubble constant by proper motions of radio components observed in AGN twin-jets.

\section{Methods and results} \label{sec.2} 
\label{sec:2}
Consider a pair of twin-jet moving bidirectionally at a relativistic velocity $\beta$ ($\beta=v/c$, where $v$ is the velocity of the ejections), with the axis of the ejections at an angle $\theta$ ($0^\circ\textless \theta\textless90^\circ$) with respect to the observer’s sightline. Under the framework of general relativity and assuming intrinsically symmetric ejections, the proper motion equation for the approaching ($\mu_a$) and receding ($\mu_r$) components can be given as  (e.g., \citealp{Rees1966,Behr1976,Blandford1979}) :
\begin{equation}
\mu_{r,a}=\frac{\beta sin\theta}{1\pm\beta cos\theta}\frac{c}{D_A(1+z)}.
\end{equation}
For sources within the Galaxy, the term of (1+z) can be omitted \citep{Mirabel1994}. From Equation (2) one finds \citep{Qin1999}: 
\begin{equation}
D_A(1+z)=\frac{c}{2\mu_a\mu_r}\sqrt{\beta^2(\mu_a+\mu_r)^2-(\mu_a-\mu_r)^2}.
\end{equation}
Combining Equations (1) and (3) yields
\begin{equation}
H_0\approx\frac{2\mu_a\mu_rz}{\sqrt{\beta^2(\mu_a+\mu_r)^2-(\mu_a-\mu_r)^2}}.\label{equ.4}
\end{equation}
From Equation (4), the law of $\beta\textless1$ leads to
\begin{equation}
H_{0,min}=z\sqrt{\mu_a\mu_r}.\label{equ.5}
\end{equation}
\begin{figure}
\centering
\includegraphics[width=0.6\columnwidth]{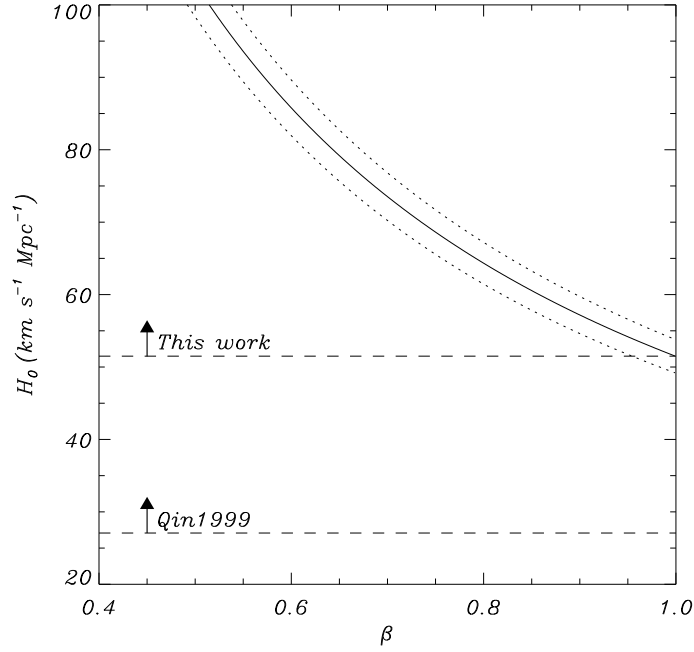}
\caption{The Hubble constant $H_0$ as a function of the real ejection velocity $\beta$ in the case of the adopted proper motions measured in the NGC 1052 twin-jet. The dotted lines indicate the uncertainty range. Two dash horizontal lines represent the lower limits of the Hubble constant obtained in \citet{Qin1999} and this work respectively.\label{fig.1}}
\end{figure}

Equation (5) shows that the lower limit of the Hubble constant ($H_{\rm 0,min}$) can be well determined, once the values of $z$, $\mu_a$ and $\mu_r$ are known. Note that $\mu_a$ and $\mu_r$ are proportional to $\beta$. For the same source, a $H_{0,min}$ calculated from the components with a larger $\beta$ value would give rise to a stronger constrain to $H_{0}$. As shown above, a distant source that is suitable for applying Equations (4) and (5) should meet the following conditions: 

(1) redshift z$\ll$1;

(2) showing twin-jet structure, and the more measurements of the proper motions of each radio components on both jet sides, the better;

(3) real velocities of certain components on both jet sides being intrinsically the same.


According to above three conditions, we find that the twin-jet in radio galaxy NGC 1052 is the most likely target source observed so. First of all, NGC 1052 is an extragalactic radio source with a redshift of 0.005037$\pm$0.000020 \citep{Denicolo2005}. Secondly, the proper motions of dozens of jet components in both sides have been intensively measured at 15 GHz \citep{Vermeulen2003,Lister2013} and 43 GHz \citep{Baczko2019}. Last but not least, bi-symmetric jet width profiles between the approaching and receding jet sides throughout scales from 300 to 4$\times$$10^7$ Schwarzschild radii have been reported \citep{Nakahara2020}, which offers strong evidence for intrinsically symmetric ejections of NGC 1052. 

The observation of the bi-symmetric jet width profiles between the two jet sides throughout so large scales strongly suggests that the motion of the jet components must be relativistic and the ejections on both sides must have almost the same strength. We accordingly assume that the ultimate ejection strengths in both sides of the twin-jet of NGC 1052 are intrinsically the same. Under this assumption, when there are enough jet components in both approaching and receding jet sides being observed, it is reasonable to consider that the real velocities of the jet components with maximum proper motions on the approaching and receding jet sides are the same. In fact, as shown in Equation (5), maximum values of $\mu_a$ and $\mu_r$ would give rise to the maximum value of the lower limit of the Hubble constant. Therefore, we adopt the maximum proper motions of the radio components in the twin-jet of NGC 1052 detected so far in our calculation, which are $\mu_a=2.156\pm0.055\rm \,mas\,yr^{-1}$ and $\mu_r=2.152\pm0.138\rm \,mas\,yr^{-1}$ for the approaching and receding components \citep{Baczko2019}, respectively. As a result, we get $H_{\rm 0,min}=51.5\pm2.3\rm \,km\,s^{-1}\,Mpc^{-1}$.

\section{Discussion} \label{sec.3}
\begin{figure}
\centering
\includegraphics[width=0.6\columnwidth]{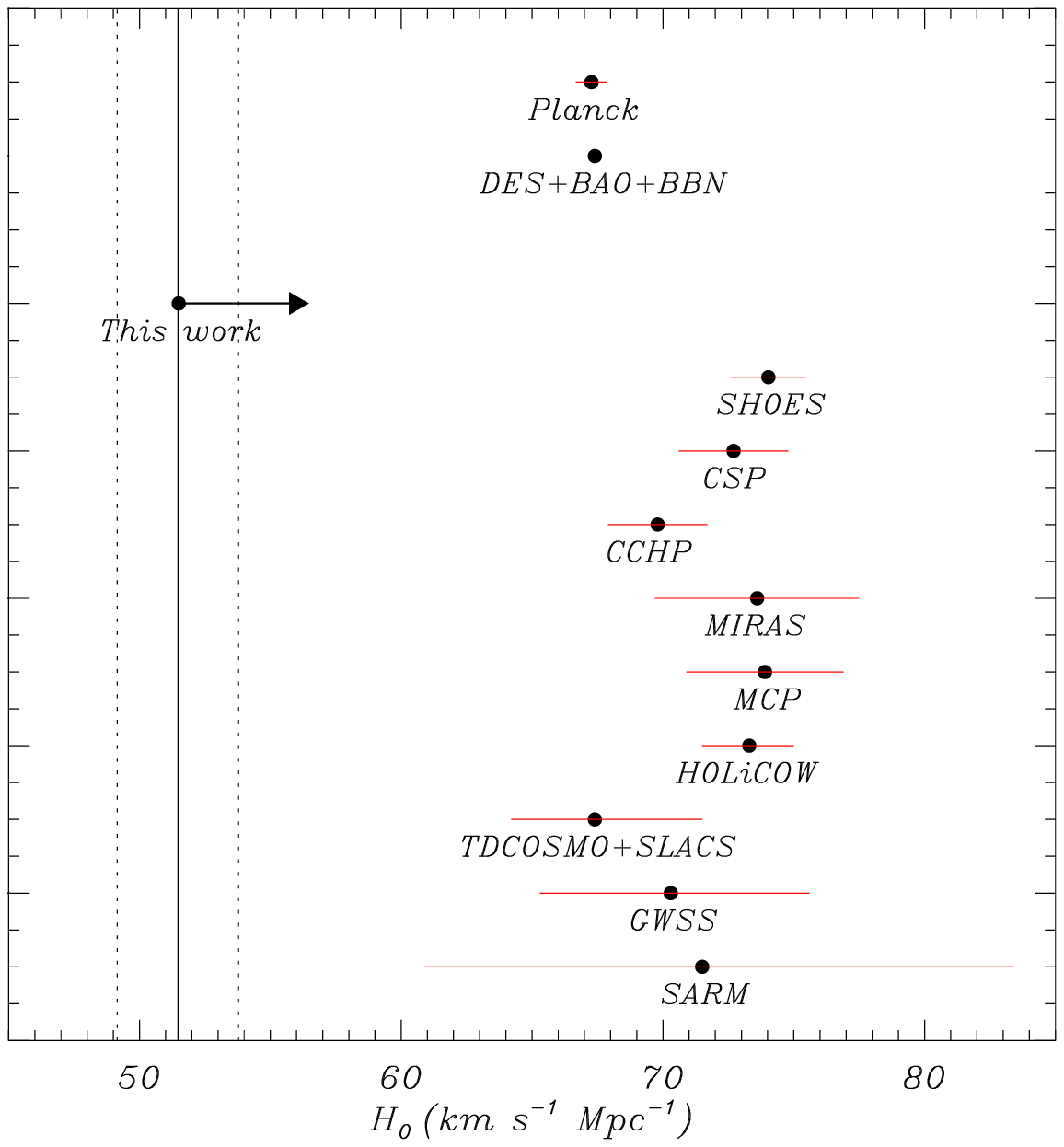}
\caption{Comparisons of various measured values of the Hubble constant. Two independent predictions of Planck \citep{Planck2020} and DES+BAO+BBN   (Dark Energy Survey+Baryon Acoustic Oscillation+Big Bang Nucleosynthesis, \citealp{Abbott2018}) are shown at the top, while the rest of the data points represent the measurements from the SH0ES  (SNe, $H_0$, for the Equation of State of dark energy, \citealp{Riess2019}), the CSP (The Carnegie Supernova Projectcite, \citealp{Burns2018}), the CCHP (The Carnegie–Chicago Hubble Project, \citealp{Freedman2019}), the MIRAS (variable red giant stars, \citealp{Huang2020}), the MCP (the Megamaser Cosmology Project, \citealp{Pesce2020}), the H0LiCOW  (The $H_0$ Lenses in COSMOGRAIL’s Wellspring Project, \citealp{Wong2020}), the TDCOSMO+SLACS \citep{Birrer2020}, the GWSS  (gravitational-wave standard siren,  \citealp{Abbott2017,Hotokezaka2019}) and the SARM  (spectroastrometry and reverberation mapping, \citealp{Wang2020}). The lower limit measurement  obtained in this work is represented by the solid vertical line and the dotted vertical lines represent the uncertainty range.\label{fig.2}}
\end{figure}

The Hubble constant as a function of the real velocity of the concerned components is well described in Equation (\ref{equ.4}). Presented in Figure \ref{fig.1} is the relation between the two quantities when adopting the above maximum proper motion values for source NGC 1052. It shows that the Hubble constant is monotonically decreasing along with the increase of the real velocity of the concerned components. Since with this method alone, one can only set the lower limit of the Hubble constant, it is reasonable to expect larger values of the lower limit of the constant in future monitoring of the twin-jet proper motions in NGC 1052. Moreover, monitoring twin-jets of other nearby extragalactic sources can also provide constraint to the Hubble constant. Among all these lower limits, the largest value should be the closest one to the real value of the Hubble constant. The more of observation, the closer to the real value is reached. Typically, with the same method, a lower limit of the Hubble constant of $H_{\rm 0,min}=27\rm \,km\,s^{-1}\,Mpc^{-1}$ was obtained in \citet{Qin1999}, while new observation of NGC 1052 in this work shows a significant larger value of the lower limit of the Hubble constant. {This is largely due to the better observation in NGC 1052. For the sources used in \citet{Qin1999}, less observation of the individual jet components was available, and for these sources, only the mean proper motions of the approaching and receding components were provided. As a contrast, NGC 1052 has been intensively observed and dozens of jet components in both sides have been measured. With these measurements, there would be a larger probability to observe jet components that have larger velocities for this source. Indeed, we are able to get a significant larger value of the lower limit of the Hubble constant from the observation of NGC 1052.}

We find that, if the real velocity of the jet component is very close to the speed of light, the values of $H_{\rm 0}$ and $H_{\rm 0,min}$ calculated from Equations (\ref{equ.4}) and (\ref{equ.5}) respectively would be almost the same. This suggests that, when more and more observations of NGC 1052 and/or other sources are available in the future, we may expect to observe jet components whose velocities are very close to the speed of light, then a certain value of $H_{\rm 0,min}$ may be fixed and no longer be exceeded. If this happens, that value would be considered a good representative of $H_{\rm 0}$. 

As illustrated above, the constraint presented in this paper does not rely on either traditional distance ladders, or the standard $\Lambda$CDM cosmological model where certain values of the cosmological parameters should be involved. It requires only well observations of proper motions of radio components observed in AGN twin-jets. Comparisons of the existing values of the Hubble constant in recent literature \citep{Planck2020,Riess2019,Abbott2018,Freedman2019,Huang2020,Pesce2020,Wong2020,Birrer2020,Abbott2017,Hotokezaka2019,Burns2018,Wang2020} and this work is presented in Figure \ref{fig.2}. It reveals that values of the Hubble constant estimated with other methods are all larger than the lower limit presented in this work. All these values are not conflicted with our constraint. We expect that future monitoring of more twin-jet sources would precent tighter boundary of the Hubble constant and probably provide contribution to solve the Hubble constant tension.

\normalem
\begin{acknowledgements}
We thank the reviewer for helpful comments. We thank Ying-Ru Lin (Baise University, China) for revising the paper. This research was supported by the National Natural Science Foundation of China (No. 11903002), and the Research Project of Baise University (No. 2019KN04).

\end{acknowledgements}
  
\bibliographystyle{raa}
\bibliography{ms2021-0194scibib}

\end{document}